# Wire GEM detector

B.M.Ovchinnikov[1], V.I.Razin, A.I.Reshetin, S.N.Filippov

Institute for Nuclear Research of Russian Academy of Sciences, Moscow

## Abstract

A wire GEM (WGEM) detector with a gas gap between meshes was constructed. The detector provides the amplification $5 \cdot 10^4$ for the gas mixture of Ar +20% $CO_2$ at atmospheric pressure. As compared with well-known GEM detectors produced by perforation the plastic plate metallized on both sides the WGEM does not suffer from breakdowns between its electrodes and the effect of accumulation of charges on holes' walls is absent. As a result the WGEM has high reliability and stability.

## Introduction

GEM detectors have good position resolution on x-, y- coordinates and may provide high resolution on z-coordinate also owing to extraction of electron components of avalanches [1]. The other advantage of GEM detectors is a high suppression of secondary avalanche production on electrodes of a chamber by photons from avalanches in GEM holes. For Penning mixture the secondary avalanches are produced after primary avalanches. The primary avalanches are produced by ionization electrons from detected event and the secondary avalanches are produced by photons from primary avalanches on component of mixture with low ionization potential [2]. Tails of secondary avalanches have continuous character and exceed considerably the primary avalanches in duration so they can be removed easily by differentiation.

Considerable shortcomings of plastic GEM are:

1. Low reliability because of accidental discharges.
2. Static charges accumulation on the walls of GEM holes leading to instability of GEM operation.

The WGEM has no such shortcomings.

## WGEM construction

WGEM electrodes were winded with Be bronze wires 0.1 mm in diameter (Fig.1). The gap between electrodes is equal to 1 mm. The holes in electrodes have dimensions 0.5 x 0.5 $mm^2$. The WGEM working area is equal to 2 cm in diameter. Schematic view of the chamber with WGEM is shown in Fig.2. Cathode electrode has a shape of a cup to ensure better charge collection to the WGEM. The signals from anode of induction gap are fed to oscilloscope DS-1080C. The chamber was filled with Ar + 20% $O_2$ mixture under atmospheric pressure. The drift gap was irradiated by α-particles having energy 4.869 MeV ($^{226}$Ra).

---

[1] Corresponding author, tel: (4967)510194, E-mail: ovchin@inr.ru

## Experimental results

The gain factor $K_{ampl}$ as a function of voltage $V_c$ is shown in Fig.3. One can see that $K_{ampl}$ is increased in range 2.3-3.0 kV up to $5 \cdot 10^4$ and thereafter become saturated. The saturation effect is caused by appearance of streamer discharges in holes of WGEM because α-particle produces in one channel of GEM $Q_e \cong 3 \cdot 10^3$ of ionization electrons and product $Q_e \cdot K_{ampl}$ (at 3kV) $\cong 3 \cdot 10^3 \cdot 5 \cdot 10^4 \cong 10^8$ is the threshold for which the streamers are appeared [3]. The streamers restrict the operating voltage range of WGEM.

These streamers in gas phase do not damage the WGEM. The operation of WGEM is restored completely after the voltage decreasing below threshold for streamer appearance.

## Conclusion

The wire GEM is constructed with electron multiplication between square holes in meshes. Because there are no plastic material between GEM electrodes the accidental discharges do not damage GEM electrodes and static charge is not accumulated between GEM electrodes. As a result the wire GEM has high reliability and stability in operation.

The wire GEM can be used in experimental physic and medicine.

**Figure captions**

Fig.1  WGEM detector layout.

Fig.2  Schematic view of the experimental setup.

Fig.3  Gain $K_{ampl}$ vs. voltage $V_c$ measured with WGEM operating in Ar + $CO_2$ using $^{226}$Ra.

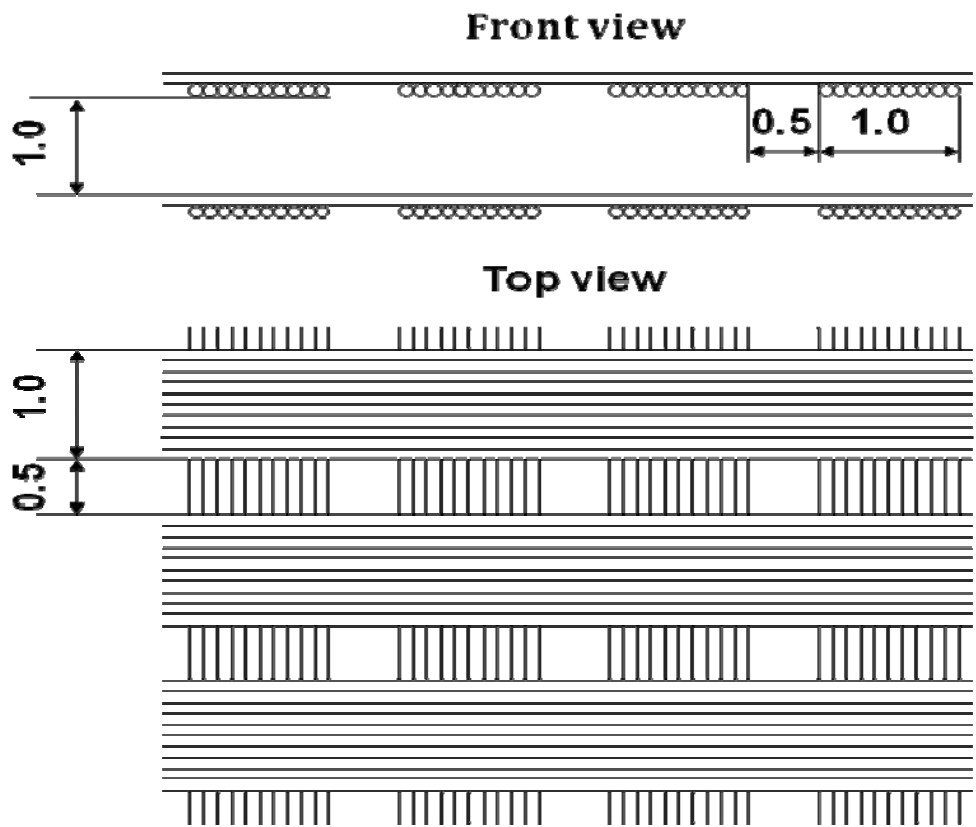

Fig. 1

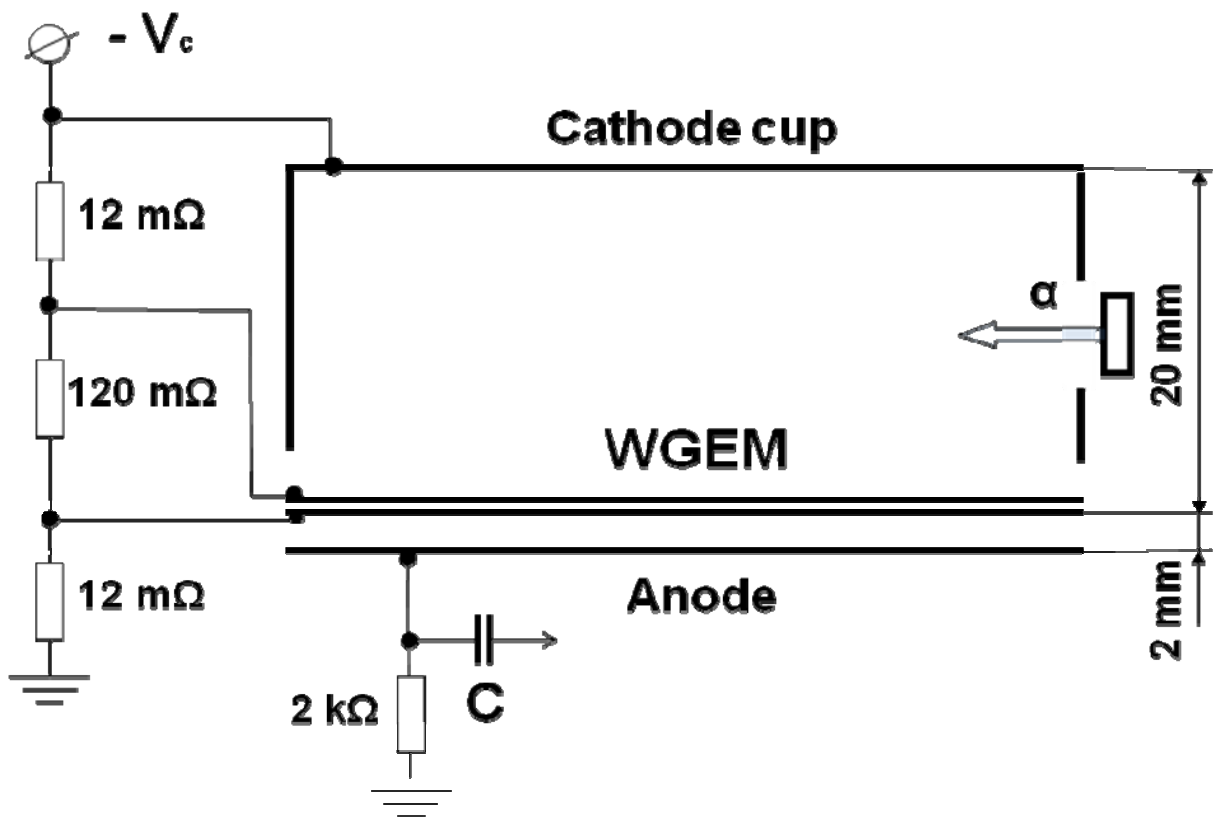

Fig.2

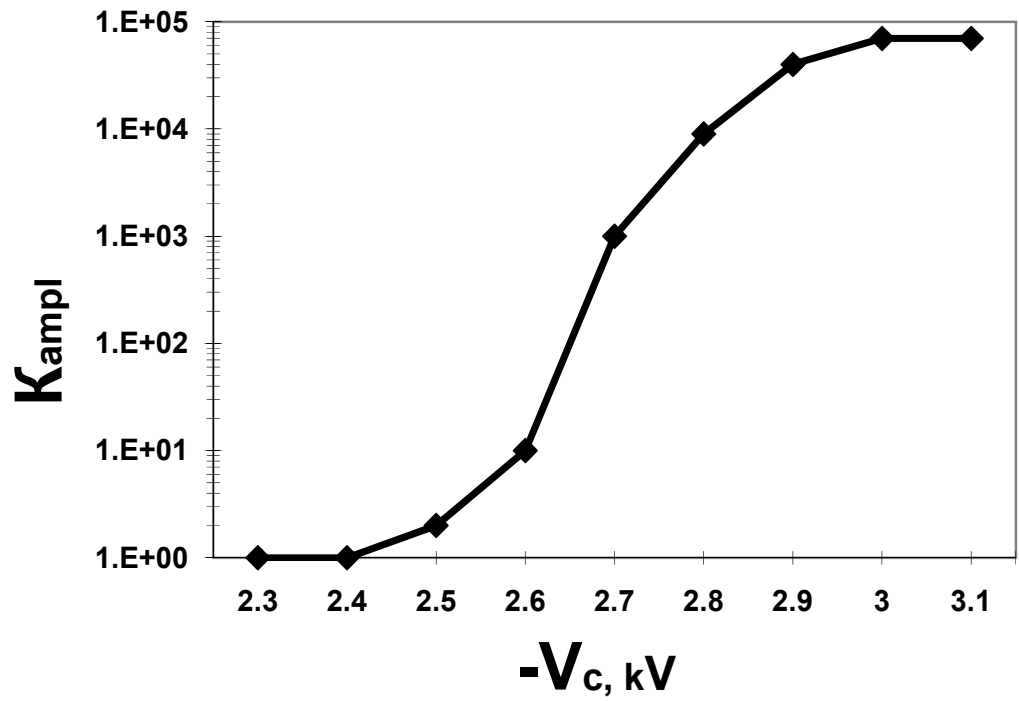

Fig. 3